\documentclass[%
reprint,
superscriptaddress,
 amsmath,amssymb,
 aps,
 pra,
]{revtex4-1}

\usepackage{graphicx}
\usepackage{dcolumn}
\usepackage{bm}
\usepackage{hyperref}
\usepackage[mathlines]{lineno}

\begin{document}


\title{The anti-proton charge radius}

\author{P.~Crivelli}
\email[]{crivelli@phys.ethz.ch}
\affiliation{Institute for Particle Physics, ETH Zurich, Switzerland}

\author{D.~Cooke}
\affiliation{Institute for Particle Physics, ETH Zurich, Switzerland}

\author{M.~W.~Heiss}
\affiliation{Institute for Particle Physics, ETH Zurich, Switzerland}

\date{\today}
\begin{abstract}

The upcoming operation of the Extra Low ENergy Antiprotons (ELENA) ring at CERN, the upgrade of the anti-proton decelerator (AD),  and the installation in the AD hall of an intense slow positron beam with an expected flux of $10^{8}$ e$^+$/s will open the possibility for new experiments with anti-hydrogen ($\bar{\text{H}}$). Here we propose a scheme to measure the Lamb shift of $\bar{\text{H}}$. For a month of data taking, we anticipate an uncertainty of 100 ppm. This will provide a test of CPT and the first determination of the anti-proton charge radius at the level of 10\%.

\end{abstract}

\pacs{Valid PACS appear here}
\maketitle


\section{\label{sec:level1}Introduction}

The measurement of the Lamb shift in atomic hydrogen is a landmark of modern physics. Lamb and Rutherford's observation of the splitting of $n=2$ states with different angular momentum \cite{lamb},  not predicted by the Dirac theory \cite{dirac},  marked the birth of quantum electrodynamics (QED). 
The most precise determination of the Lamb shift using microwaves to induce directly 2S$\to$2P transitions was performed by Lundeen and Pipkin who measured a value of 1057.845(9) MHz \cite{lundeen}. Making use of Ramsey's separate oscillating field (SOF) technique \cite{ramsey} they were able to reduce the 100 MHz natural linewidth, which is limited by the radiative lifetime of the 2P state, down to a width of 30 MHz.

 The finite size of the proton contributes with a correction that is given by \cite{LSRProton}:
\begin{equation}
\Delta E=\frac{1}{12}\alpha^4 m_r^3 r_p^2
\end{equation}
where $\alpha$ and $m_r$ are the fine structure constant and the hydrogen reduced mass, thus, from the Lamb shift determination one can extract the proton charge radius  $r_p$ at a level of 3\% \cite{savely}.  Motivated by the proton radius puzzle prompted by the muonic hydrogen experiment at PSI \cite{MuH}, new efforts led by E. A. Hessels have been undertaken at the York University in Toronto to improve the precision of the Lamb shift. With their clever refinement of the SOF technique, E. A. Hessels et al. should be able to reduce the systematic in order to determine $r_p$ at a level of 1$\%$ uncertainty \cite{Hessels}. 

Here we propose a way to measure the Lamb shift of anti-hydrogen.  Such a measurement was proposed in 1998 at Fermilab using relativistic  $\bar{\text{H}}$ and should have resulted in an uncertainty of 5 \% \cite{blandford}. Our proposed scheme relies on a completely different technique which should lead to almost two orders of magnitude higher precision resulting in a stringent test of CPT and the first determination of the anti-proton charge radius. 

 Anti-hydrogen is a blossoming field of research which studies aim to shed light on the observed baryon asymmetry in the Universe. In addition to the origin of Dark Matter, this is probably one of the most tantalising puzzles of modern particle physics and cosmology that seeks for an answer. 

\begin{figure}[h!]
\centering
\includegraphics[width=.45\textwidth]{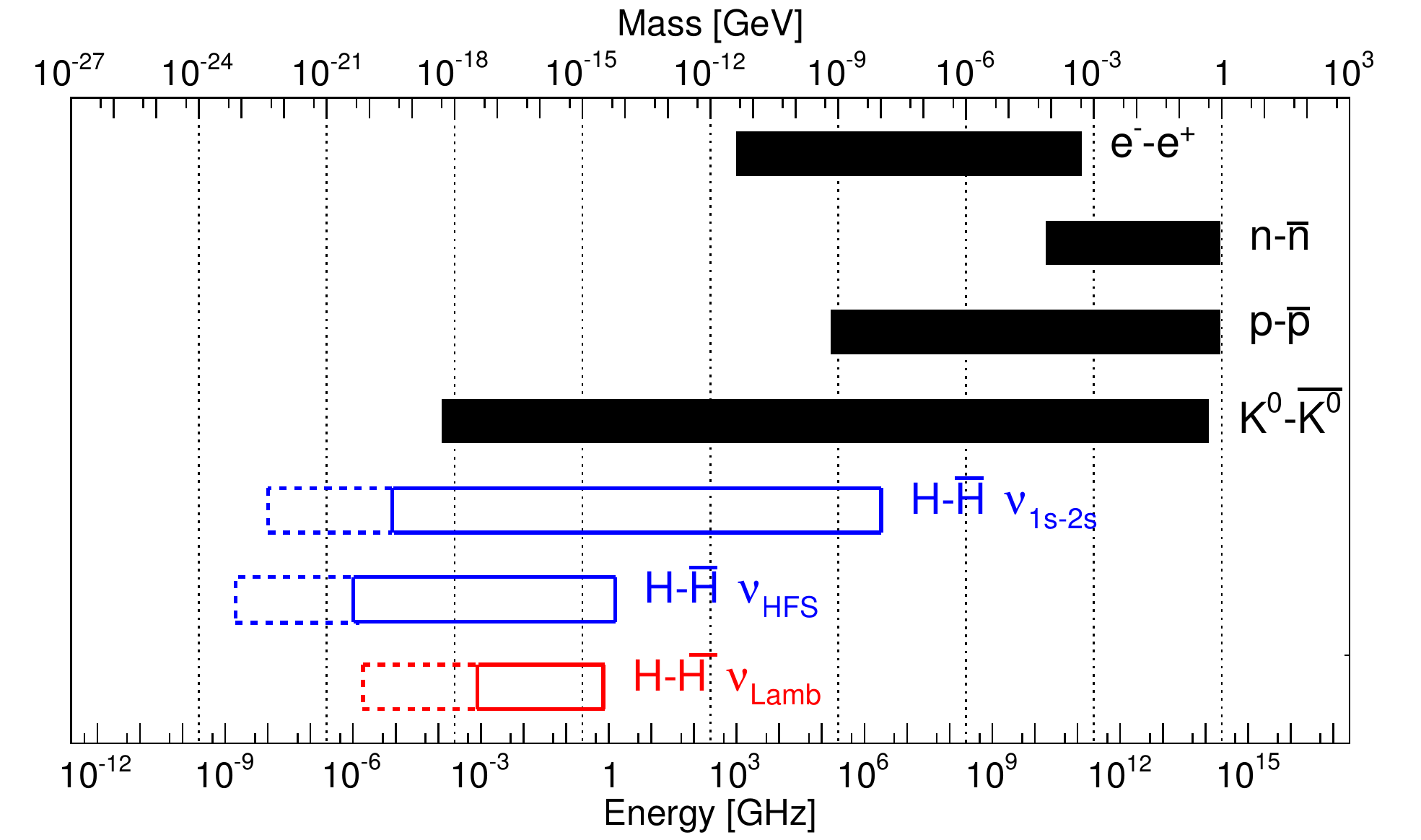}
\caption{Sensitivity as a test of CPT for different systems and projected accuracy for anti-hydrogen experiments including the Lamb shift measurement proposed here (solid line) and sensitivity if a relative accuracy as for the matter counterpart could be achieved (dotted line). This Figure was adapted from \cite{Eberhard}.}
\label{fig:CPT_scheme}
\end{figure}

The first observation of $\bar{\text{H}}$ at the CERN Low Energy Anti-protons Ring (LEAR) \cite{LEARHBAR}, motivated the construction of the Anti-proton Decelerator (AD) facility which allowed the production of antihydrogen ($\bar{\text{H}}$) at low energies \cite{ATHENA2002, ATRAP2002}. The formation of $\bar{\text{H}}$ was achieved by mixing trapped positrons and antiprotons plasmas in a nested Penning--Malmberg trap \cite{NestedTrap}. With refinements of this technique, $\bar{\text{H}}$ can now be magnetically trapped \cite{ALPHA2010,ALPHA2011, ATRAP2012}. This important milestone will allow for precision laser spectroscopy and thus promise to provide a test of CPT to a very high accuracy. An alternative method to form a $\bar{\text{H}}$ beam to be used for a precise hyperfine splitting measurement \cite{ASACUSA2014HFS}, was recently demonstrated making use of a CUSP trap \cite{ASACUSA2010}. The projected accuracy of these measurements and the one proposed here are summarized in Fig. \ref{fig:CPT_scheme} together with the existing CPT limits parametrised in terms of the absolute accuracy which can be used for comparison of different systems \cite{Kostelecky}.

Anti-hydrogen is also being used to test the gravitational behaviour of antimatter. A first direct limit has been inferred on the gravitational acceleration of antimatter by releasing the $\bar{\text{H}}$ atoms from the magnetic trap \cite{ALPHAGBAR}. Improvements on this setup, comprising laser cooling or a vertical magnetic trap \cite{ALPHAGBARPlus}, could lead to a test of the effect of gravity on antimatter with an accuracy of 1\% or better. With the same goal two proposals have been approved at CERN \cite{Aegis, GBar}. Both experiments are ongoing and they are planning to produce anti-hydrogen or anti-hydrogen ions via charge exchange of positronium (Ps) with anti-protons:\\
\begin{eqnarray}
& \text{Ps}+\bar{\text{p}}\to\bar{\text{H}}+\text{e}^-\\
& \text{Ps}+\bar{\text{H}}\to\bar{\text{H}}^++\text{e}^-.
\end{eqnarray}
Production of normal hydrogen in its ground state via charge exchange of protons with positronium has been demonstrated by M. Charlton et al. \cite{Charlton1997}.
The same mechanism has already been proven to produce $\overline{\text{H}}$ in Rydberg states ($\overline{\text{H}}^*$) by the ATRAP collaboration using a two step charge exchange, i.e. formation of Rydberg positronium with positron on Cs atoms and subsequent formation of $\overline{\text{H}}^*$ \cite{ATRAP2004,ATRAPJMol1_2016}. 
The cross sections for the charge exchange reactions were calculated by different authors \cite{Mitroy1995}-\cite{Bailey2015} and good agreement has been found with the available experimental data. Recently, new improved calculations were performed \cite{Charlton2016}.

\section{Proposed scheme}

\begin{figure}[h!]
\centering
\includegraphics[width=0.45\textwidth]{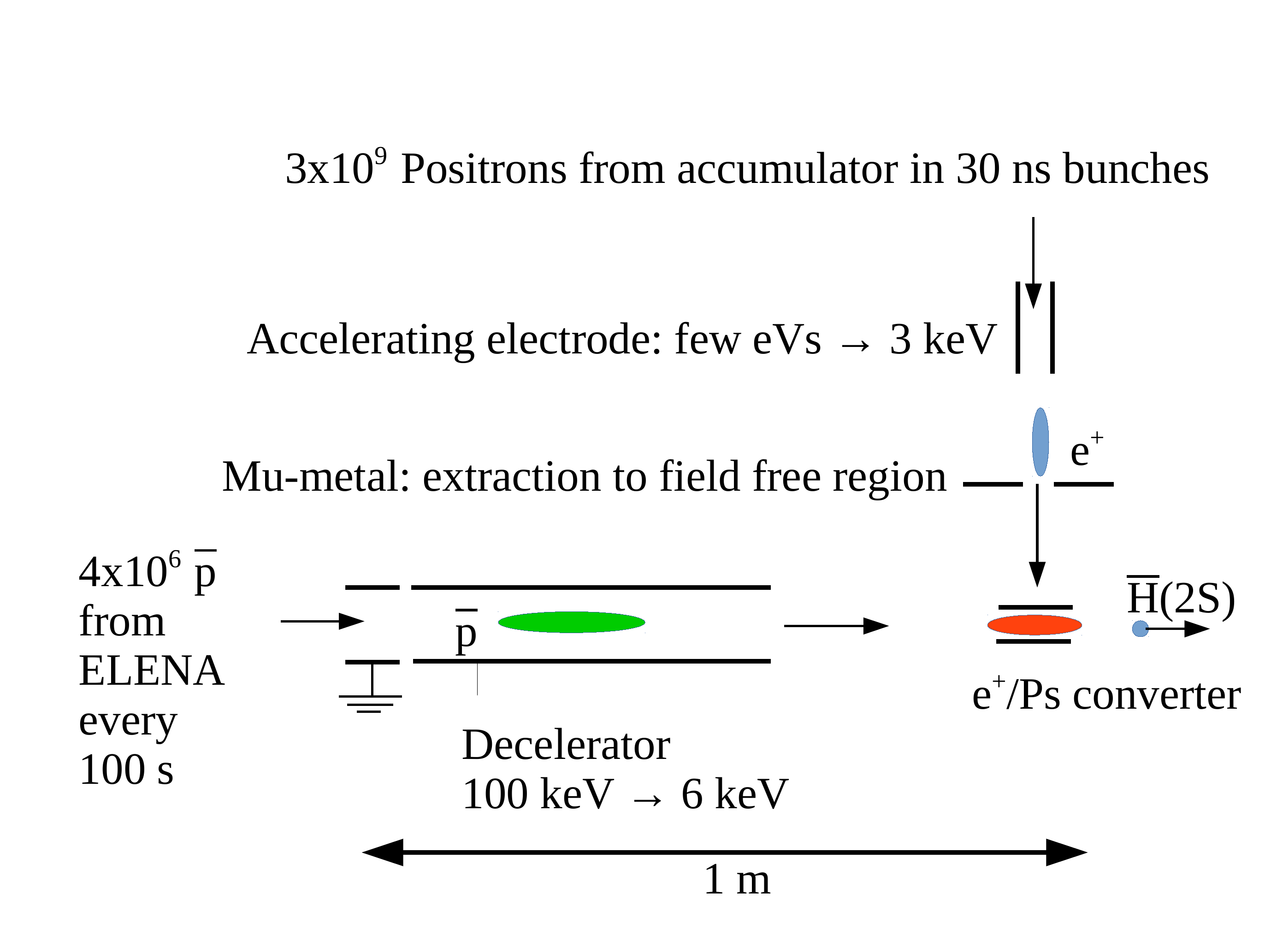}
\caption{Scheme for the production of the metastable 2S $\bar{\text{H}}$ beam.}
\label{fig:LSSchemeProd}
\end{figure}

\begin{figure}[h!]
\centering
\includegraphics[width=0.45\textwidth]{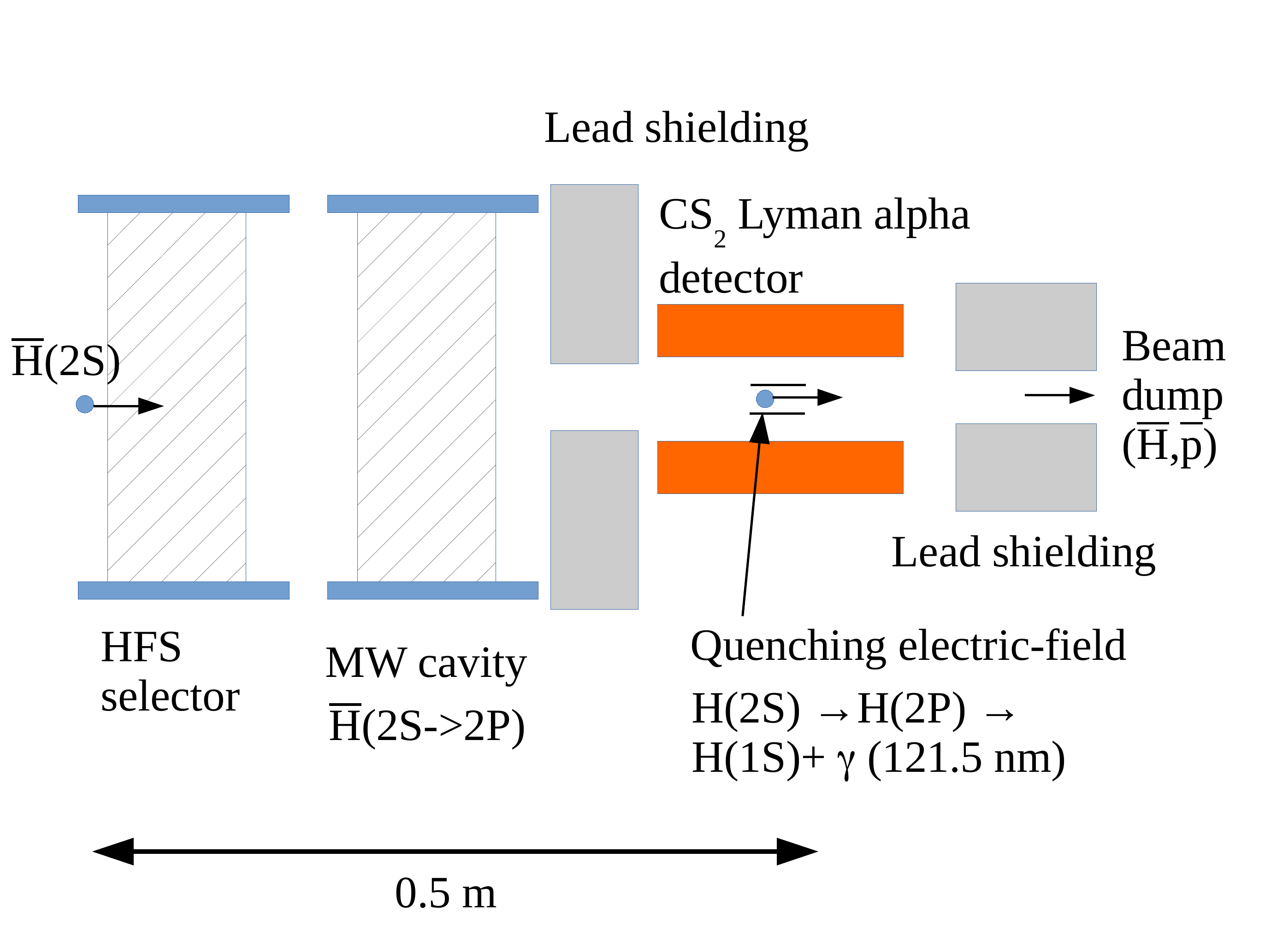}
\caption{Scheme of the proposed experimental setup for the measurement of the $\bar{\text{H}}$ Lamb shift.}
\label{fig:LSSchemeDet}
\end{figure}

In the scheme proposed here, a metastable 2S $\bar{\text{H}}$ beam is produced via charge exchange reaction between the anti-protons and a dense positronium (see Fig. \ref{fig:LSSchemeProd}).  
The 2S atoms will pass through two microwave (MW) field regions where unwanted hyperfine states are removed by a state selector and the 2S$\to$2P transitions are subsequently induced (see Fig. \ref{fig:LSSchemeDet}). After this region, a static electric field quenches the remaining 2S atoms in the 2P states that de-excite in 1.6 ns to the ground state via emission of the Lyman alpha photon. The detection of these photons allows to measure the quenched fraction as a function of the MW frequency and thus to determine the $\bar{\text{H}}$ Lamb shift as was done for its matter counterpart \cite{Newton1979}.
Positronium is formed implanting keV positrons in a thin SiO$_2$ porous films \cite{APL}. At 3 keV positron implantation energy, the conversion efficiency for Ps emitted into vacuum with an energy of about 80 meV is 35\% \cite{oPsTOF,CassidyPRA}. The electrons from the 10 MeV LINAC (under installation in the AD by the GBAR collaboration) impinging on a tungsten target will produce positron-electron pairs \cite{OurLINAC}. The positrons are moderated with tungsten meshes resulting in a slow positron beam flux of $3\times10^8$ e$^+$/s that will be injected in a buffer gas trap (typical efficiency of 20 \% \cite{Surko}) and subsequently  transferred to a 5 T accumulator \cite{Oshima}. In 100 s (the AD cycle), $3\times10^9$ positrons can be stacked and cooled via cyclotron radiation in the accumulator. Note that accumulation of $4\times10^9$ positrons has recently been demonstrated \cite{ATRAPJMol2_2016}. By lowering the last electrode of the 5 T trap the positrons are released in bunches of 30 ns that will be synchronised with the $\bar{\text{p}}$ pulse. The diameter of the positron beam at the exit of the trap is around 60 $\mu$m \cite{Surko}. The positrons will be guided in a decreasing magnetic field until this reaches 100 G. The beam diameter will be of the order of $r_{100G}=r_{5T}\sqrt{\frac{5\text{T}}{10^{-3}\text{T}}}=1$ mm \cite{Surko}. 
At this point they can be injected in a drift tube to accelerate them to 3 keV by applying an HV pulse when the positrons are inside the tube \cite{Cooke2015b}. From a 100 G field the positrons will be extracted to an electromagnetic field-free region, as demonstrated with 90\% efficiency \cite{Cooke2015b}, and refocussed to a 1 mm beamspot. This method has the advantage that the target is at ground potential and thus electric fields can be avoided when extracting the 2S $\bar{\text{H}}$ beam which ensures that no quenching to the 2P state will occur, and magnetic fields perpendicular to antiproton propagation are avoided. 

The 3 keV positrons will enter the target passing through a thin (30 nm) SiN window coated with a 3 nm gold layer (to avoid charging) with an efficiency close to 100\% \cite{crivelli2014}. At this energy, the Ps formed in the silica film takes about 1 ns to be emitted into the tube \cite{CassidyDelayed}. The mean implantation depth is 100 nm resulting in an instantaneous Ps density of $4\times 10^{11}$ Ps/cm$^3$. Using the measured cross sections for spin-spin quenching between Ps with $m=1$ and $m=-1$ and for Ps$_2$ formation  \cite{DensePs}, one can estimate that the losses through this mechanism are negligible.

The Ps formed in the target will then be confined in a tube allowing to keep a high Ps density. This geometry was used in the 1S-2S Ps experiment at ETH Zurich to observe 2S annihilations \cite{hype15} and it is planned to be implemented in the GBAR experiment \cite{GBar}.
A Monte-Carlo simulation in Geant4 \cite{geant4}, validated by our experiments with Ps (see e.g. \cite{oPsTOF,oPsMOF,hype15}), is used to calculate the time evolution of the Ps density in the tube. 

The ELENA ring is expected to deliver $4.5\times10^6$ $\bar{\text{p}}$ with 100 keV momentum in 75 ns bunches every 100 s \cite{ELENA}.
Since the maximum of the cross section for  $\bar{\text{H}}$ production in the 2S state is at 6 keV \cite{Charlton2016} a lower momentum of the anti-protons is preferable.
This could be achieved using the scheme proposed by D. Lunney for the GBAR experiment \cite{LunneyGBAR}. A drift tube kept at -94 kV is pulsed down to ground while the $\overline{p}$ bunch is inside it. The antiprotons will thus find themselves at the exit of the buncher tube at the potential given by the difference of their initial energy with the HV pulsed on to the tube, i.e. 6 kV. Such a technique is routinely used in ISOLDE to decelerate ions \cite{ISOLDE}.
The deceleration results in an increase of the beam emittance which can be estimated from the input and output energies as: $\epsilon^{6\text{keV}}=\epsilon^{100\text{keV}}\sqrt{100/6}=16 \pi $ mrad mm where  $\epsilon^{100\text{keV}}=4\pi$ mrad mm is the expected emittance of ELENA. SIMION simulations confirm that this estimate is a good approximation. Focussing the antiprotons to optimally fit through the 20 mm long $1\times1$ mm$^2$ Ps formation tube results in a $\sim50$\% geometrical transmission factor with a beam spot of 0.8 mm diameter. 
	 
\begin{figure}[h!]
\centering
\includegraphics[width=0.45\textwidth]{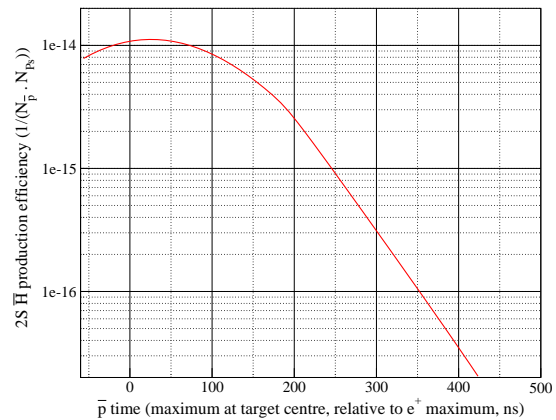}
\caption{Calculated $\overline{\text{H}}$(2S) atom yield per number of Ps and antiprotons.}
\label{fig:production2S}
\end{figure}

Using the calculated cross section and the simulated time evolution of the Ps density the probability per 6 keV antiprotons and Ps(1S) for $\overline{\text{H}}$(2S) atom yield is calculated as shown in Fig. \ref{fig:production2S}. The energy and direction of the $\overline{\text{H}}$ atoms is defined by the initial antiprotons momentum, thus a beam with an emittance of 16 $\pi$ mrad mm can be formed resulting in a beam size of the order of 4 cm at a 0.5 m distance. 
The expected flux of $\overline{\text{H}}$(2S) atoms per AD cycle is given by:
 \begin{equation}
 N_{\overline{\text{H}}{\text(2S)}}=\epsilon_{ce}  \cdot \epsilon_{dt}\cdot N_{\overline{\text{p}}} \cdot \epsilon_{em} \cdot  \epsilon_{dt}\cdot N_{e^+}
\end{equation} 
where $\epsilon_{ce}=10^{-14}$ is the charge exchange yield taken from Fig. \ref{fig:production2S}, $\epsilon_{dt}=0.5$ is the expected transmission of decelerated antiproton pulse ($N_{\overline{\text{p}}}=4.5\times10^6$)  through the Ps formation tube, $\epsilon_{Ps}=0.35$ is the conversion efficiency of positrons to Ps emitted into vacuum and $N_{e^+}=3\times10^9$ is the number of accumulated positrons extracted to a field free magnetic region with an efficiency of $\epsilon_{em}=0.9$.
A flux of about $N_{\overline{\text{H}}\text{(2S)}}=20$ per AD cycle is anticipated of which 25\% will be in the F=0 state.

\begin{figure}[h!]
\centering
\includegraphics[width=0.45\textwidth]{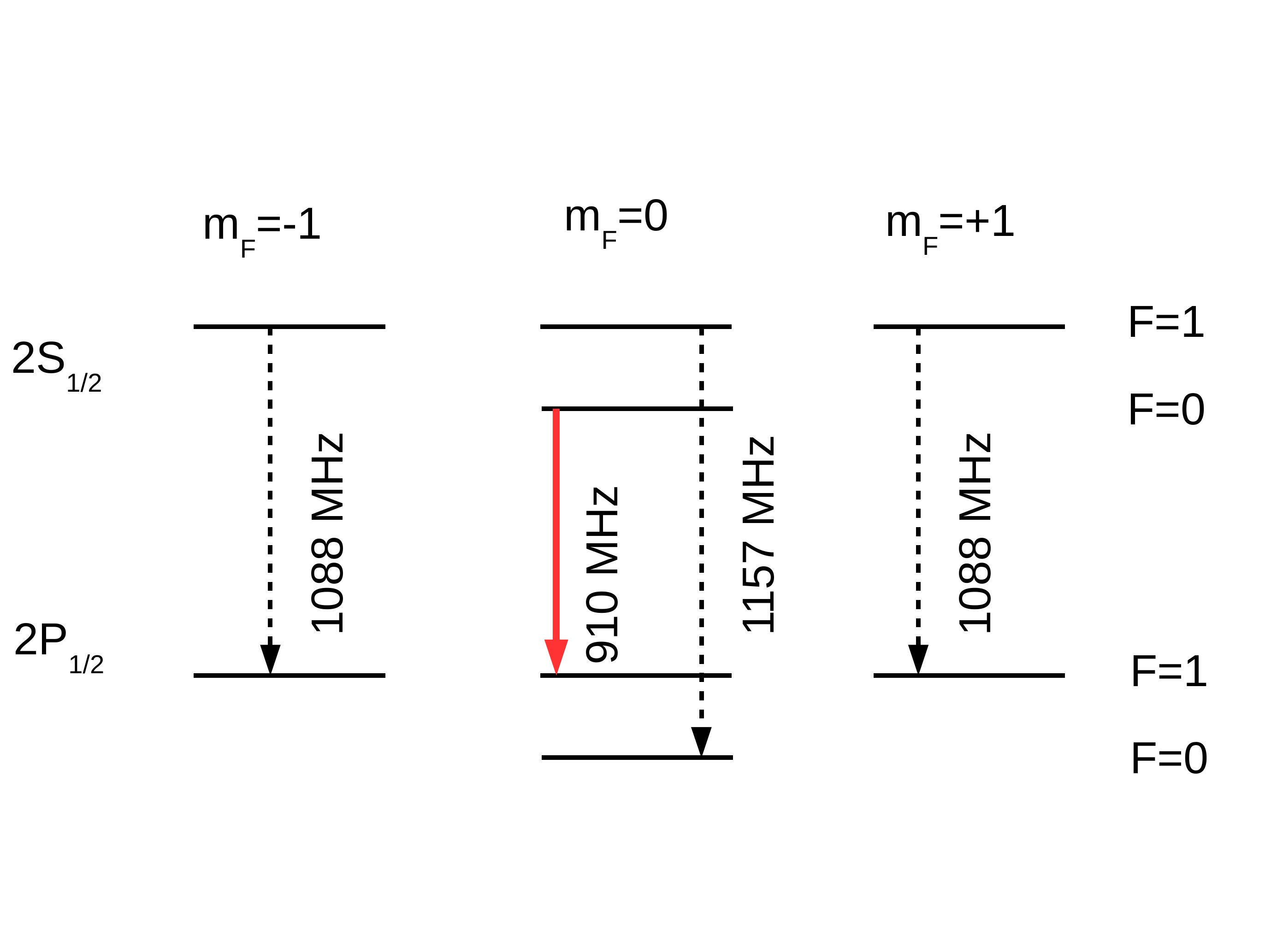}
\caption{Sketch of the 2S and 2P hyperfine levels.}
\label{fig:hf2S}
\end{figure}

The simplified scheme of the 2S and 2P hyperfine levels is shown in Fig. \ref{fig:hf2S}.
In order to isolate the 2S$_{1/2}$ F=0 transition, an hyperfine state selector will be used. 
Newton et al. \cite{Newton1979} used a single frequency to drive down all the 2S$_{1/2}$, F=1 states to the short lived 2P$_{1/2}$ retaining 20\% of the F=0 population. With a higher Q cavity (loaded Q of 2500) tuned at 1139 MHz, we estimated that the F=1 population can be quenched below 1\% while 60\% of the F=0 population are retained. 


Note that for 6 keV energy of the antiprotons and for ground state Ps the probability to form $\overline{\text{H}}$ in higher states is very suppressed \cite{Charlton2016}. We calculated that the number of atoms cascading down to the 2S state after the HFS state selector is at negligible level. 

The 2S, F=0 atoms will pass through a single microwave region where the 2S$_{1/2}\to$2P$_{1/2}$ F=0 transitions can be induced with high efficiency by tailoring the MW power to time of flight of the atoms in the MW field in order to achieve a $\pi$ pulse. This can be achieved by using a cavity with a loaded Q value as low as 400 for reasonably low MW input powers.

Even though the SOF technique would allow for a reduction of the natural line width by a factor 3-5, it has the drawback of greatly reducing the signal rate. Therefore,  due to the small number of  $\bar{\text{H}}$ atoms available, a single MW region is preferable (at least as a first step).

An electric field will be used to quench the surviving 2S to the 2P states. The emitted Lyman-alpha photons will be detected in a cylindrical CS$_2$ gas photo-ionization detector \cite{CS2, bollinger1981}.
Efficiencies higher than 50\% including the losses on the LiF window were reported for this kind of  detectors which can be arranged in such a way to achieve a solid angle coverage close to 4$\pi$.
Pulsed beam operation combined with time of flight will result in an excellent S/N ratio. The width of the signal window will be about 1 $\mu$s for which the accidental rate is estimated to  be at a level of $10^{-3}$.

%
%

\section{Expected accuracy}
The estimated signal on resonance per bunch is given by :
\begin{equation}
N_d=\epsilon_q  \cdot \epsilon_t \cdot   \epsilon_d \cdot N_{\bar{\text{H}}\text{(2S)}}
\end{equation}
where 
$\epsilon_q=0.6$ is the surviving fraction of $N_{\bar{\text{H}}(2S)}$ with F=0 in the hyperfine state selector, $\epsilon_t=1$ is the transition probability 2S$\to$2P, $\epsilon_d=0.5$ is the probability for the detection of the Lyman alpha  photon, $N_{\bar{\text{H}}(2S)}$ are the number of atoms produced in 2S$_{1/2}$ state with F=0.

The number of events detected per day on resonance assuming a duty cycle of ELENA of 80\% will be around 1000 events. Simulation of the expected line shape predicts that with the expected S/N ration a month of data taking the line centre can be determined with an uncertainty of 100 ppm. 

The main source of systematic is the AC Stark shift which will be below 100 kHz. By measuring the line shape at different MW powers this could be corrected for extrapolating to zero intensity. 
Other sources of systematic are the first and second order Doppler shift which for the given momentum and spread after the decelerator of the ${\bar{\text{H}}(2S)}$ are at a level of 10 kHz. Other shifts such as motional Stark Shift and Zeeman at a level of few kHz, assuming the magnetic field in the excitation region will not exceed the field of the earth.

\section{Conclusions}
A scheme to measure the anti-hydrogen Lamb Shift at a level of 100 ppm has been proposed. Such a measurement might be feasible in the near future thanks to the ELENA ring, the ongoing upgrade of the AD, and the installation of an intense positron source based on a 10 MeV LINAC at CERN.
This experiment would result in a stringent test of CPT and the first determination of the anti-proton charge radius at a level of 10\%.  The accuracy in the experiment will limited by statistics. If improvements in the production rate of meta-stable 2S anti-hydrogen atoms could be achieved, it would be conceivable to reach a precision of few ppm close to the one reached for its matter counterpart.

\begin{acknowledgments}
The authors gratefully acknowledge P. Comini for the very useful discussions on the charge exchange process and A. Antognini and F. Nez for their very valuable comments and suggestions. 
This work has been supported by the Swiss National Science Foundation under the grant number 200020\_166286.

\end{acknowledgments}

\end{document}